\documentclass[12pt]{article}
\usepackage{graphicx}

\begin{document}

\hoffset = -1truecm \voffset = -2truecm \baselineskip = 10 mm

\title{\bf Partonic structure of proton in the resonance region}

\author{{\bf Jianhong Ruan}$^a$, {\bf Rong Wang}$^b$, {\bf Xurong Chen}$^b$ and {\bf Wei Zhu}$^a$,
\\
\normalsize $^a$Department of Physics, East China Normal University,
Shanghai 200062, P.R. China \\
\normalsize $^b$Institute of Modern Physics, Chinese Academy of
Sciences, Lanzhou 730000, P.R. China\\
}

\date{}

\newpage

\maketitle

\vskip 3truecm

\begin{abstract}

We separate the contributions of parton distributions from higher twist corrections
to the deeply inelastic lepton-proton scattering in the resonance region
using the Jefferson Lab data at low $Q^2$. The study indicates that
the concept of the valence quarks and their distributions are indispensable
even at $Q^2<1GeV^2$. The quark-hadron duality is also discussed.

\end{abstract}

PACS number(s): 25.30.Dh, 25.30.Fj, 13.60.Hb

$keywords$: Structure function; Parton model; Higher twist; Quark-hadron duality \\

\newpage

The theoretical description of nucleon structure in the transition region
between perturbative and non-perturbative QCD is still an open question.
Structure function data in the lepton-nucleon deeply
inelastic scattering (DIS) are important measurements in the
investigations of the nucleon structure. Jefferson Lab (JLab) has
measured the proton structure function $F_2^p$ at $0.07\leq
Q^2\leq 3.3 GeV^2$ and $W^2<4GeV^2$ [1] ($W^2$ is the invariant
mass of the final hadronic state) almost two decades ago.
However, we still do not have a complete understanding of this experiment.

According to standard model, proton consists of quarks and gluons, which are
complicatedly correlated to each other by strong interaction.
In this work we try to extract the information about the quark structure
in the proton at low $Q^2$ (down to $0.07GeV^2$) from the JLab data.
According to the operator product expansion (OPE) method,
the proton structure function can be decomposed as
$$F_2^p(x,Q^2)=F_2^{LT}(x,Q^2)+F_2^{HT}(x,Q^2),\eqno(1)$$
where $F_2^{LT}(x,Q^2)$ is the leading twist contribution, and
$F_2^{HT}(x,Q^2)$ is the higher twist corrections.
The leading twist (LT) part of the structure function directly relates to
the parton distributions, which are usually defined as the single,
non-interacting partons at $Q^2>1GeV^2$ and their $Q^2$-evolution
is well studied by pQCD theory. However the quantity of higher twist (HT) corrections
is not clear in QCD theory. Because it corresponds to multi-parton interactions and
connected to the complicated nonperturbative QCD effects.
It is expected that the structure functions at low $Q^2$ can provide
important information of the higher twist effects since they are
power $1/Q^{2n}$-dependent.

To separate the higher twist effects using the JLab data at low $Q^2$ in Ref. [1],
we need to know: (a) how to measure $F_2^p(x,Q^2)$ in the resonance region?
(b) how to calculate the $F_2^{LT}(x,Q^2)$ at $Q^2<1GeV^2$?
For the first question, the work [1] has obtained the single curves
$\overline{F}_2^p(x,Q^2)$ using the quark-hadron duality [2],
which are the integrals of the resonance curves integrated in a defined $x$-range,
corresponding to the resonance region. In this work we take
$$F_2^p(x,Q^2)=\overline{F}_2^p(x,Q^2),\eqno(2)$$
in the resonance region from Ref. [1]. For the second question,
unfortunately, most of the PDF global analyses are performed in
the range of $Q^2>1GeV^2$. There are only two global analyses
which give the parton distributions at low $Q^2<1GeV^2$. One of
them is the Gl$\ddot{u}$ck-Reya-Vogt (GRV) PDFs [3], where the
valence quarks and valence-like gluon evolve from $\mu^2\sim
0.3GeV^2$ according to the DGLAP equation [4]. The other is PDFs
from our previous work[5][6]: pure valence quark distributions
starting from $\mu^2=0.064GeV^2$ evolve according to a modified
DGLAP (the GLR-MQ-ZRS) equation [7], which modifies the evolution
of parton distributions with recombination effects but not
included $F_2^{HT}$ in Eq. (1). We find that the negative
nonlinear corrections in the GLR-MQ-ZRS equations improve the
perturbative stability of the QCD evolution equation at low $Q^2$.
The resulting parton distributions of proton at the leading order
approximation with only four free parameters are consistent with
the existing data. The nonlinear equation provides a powerful tool
to connect the non-perturbative quark model with the measured
structure functions.

The single curves in Ref. [1] matched the
valence quark distributions. It seems that the valence quark
distributions and even a pQCD analysis of them are available and
reliable (at least partly) at $Q^2<1GeV^2$. Therefore, as a model,
we try to use the predicted $F_2^{ZRS}(x,Q^2)$ in [5] to be
$F_2^{LT}(x,Q^2)$ in the valence quark kinematic range, i.e.,
$$\overline{F}_2^p(\xi,Q^2)=F_2^{ZRS+TMC}(\xi,Q^2)+F_2^{HT}(\xi,Q^2)$$
$$=\left [1+\frac{C_{HT}(\xi,Q^2)}{Q^2}\right ]F_2^{ZRS+TMC}(\xi,Q^2), \eqno(3)$$
where $C_{HT}$ is the effective higher twist coefficient. The
target mass corrections (TMC) take into account the finite mass
effect of the initial nucleon and
$\xi=2x/(1+\sqrt{1+4x^2m^2_p/Q^2})$ is the Nachtmann variable [8].

\begin{figure}[htp]
\begin{center}
\includegraphics[width=0.5\textwidth]{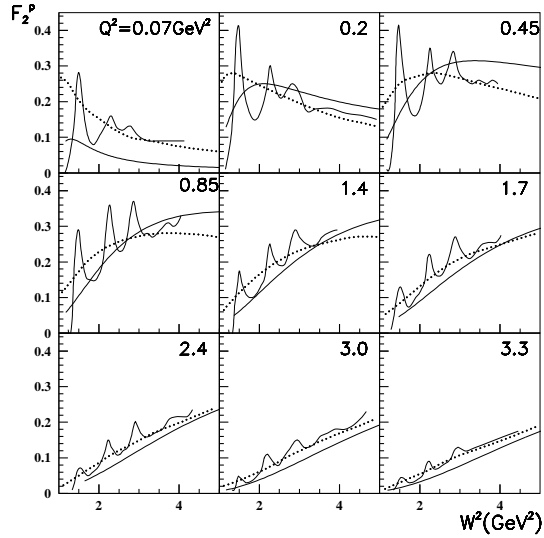}
\caption{
Proton structure functions as the function of the invariant
mass squared $W^2$ in the nucleon resonance region. Fluctuating
curves: measured data by JLab [1]; Dotted curves: the scaling
curves using quark-hadron duality [2]; Solid curves: Our predicted
$F_2^{ZRS+TMC}$.
}
\label{fig:1}
\end{center}
\end{figure}

\begin{figure}[htp]
\begin{center}
\includegraphics[width=0.5\textwidth]{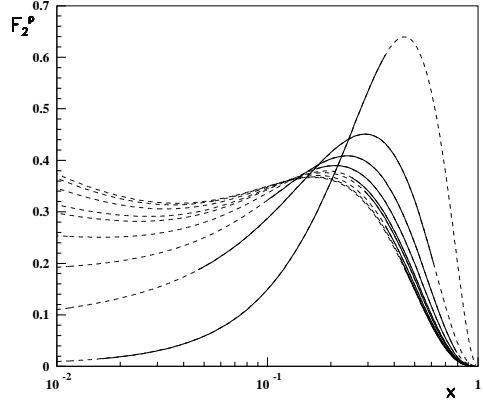}
\caption{
The pQCD evolution of $F_2^{ZRS}$ from top $Q^2=0.07GeV^2$ to
$3.3GeV^2$. The scaling violation is obvious at the beginning
evolution at $Q^2<1GeV^2$.
}
\label{fig:2}
\end{center}
\end{figure}

The comparisons of $F_2^{ZRS+TMC}$ with
$\overline{F}_2$ are presented in Fig. 1.
$F_2^{ZRS}(x,Q^2)$ are shown in Fig. 2, where the solid curves show the
measured kinematic range of JLab data. One can find that $F_2^{ZRS}(x,Q^2)$
and $\overline{F}_2^p(x,Q^2)$ have similar shapes on both
sides of the valence quark peak. This similarity between $F_2^{ZRS}(x,Q^2)$
and $\overline{F}_2^p(x,Q^2)$ strongly implies that the
valence quark distribution and its pQCD evolution are indispensable
even at very low $Q^2$.

The following work is to extract the higher twist coefficients
taking the values of the JLab-single curves measurement [1] and
$F_2^{ZRS+TMC}$ from the global analysis [5] by using Eq. (3). The
obtained $C_{HT}(\xi,Q^2)$ are shown in Fig. 3, where the last
three plots at high $Q^2$ are taken from [9]. What is the physical
interpretation of the obtained $C_{HT}(\xi,Q^2)$ in the resonance
region? Power corrections originate from the multi-parton
interactions at the final state. Figure 3 shows that $C_{HT}\sim
0$ at $Q^2=0.07GeV^2$, which is near the starting scale of the
pQCD evolution in our previous work. A possible explanation is
that the proton mainly consists of three valence quarks and all
non-perturbative effects are absorbed into the definition of the
initial valence quark distributions, which consequence in
$C_{HT}=0$ at $Q^2=0.064GeV^2$. This small $C_{HT}$ at very low
$Q^2$ coincides with a smooth transition to the expected behavior
in the real photon scattering limit $F_2(\xi,Q^2)\rightarrow 0$
for $Q^2\rightarrow 0$, which requires
$$C_{HT}(\xi,Q^2)\rightarrow -Q^2, at~Q^2\rightarrow 0.\eqno(4)$$
We can guess that the higher twist coefficient gradually increases
with the creation of new partons in pQCD evolution. However, at
the same time, some of the correlations among the partons
disappear under an impulse of the probe with high $Q^2$-the
impulse approximation [10]. Under the influence of the above two
opposite processes, the higher twist coefficient reaches a balance
in the range of $Q^2\sim 1-3 GeV^2$.

\begin{figure}[htp]
\begin{center}
\includegraphics[width=0.5\textwidth]{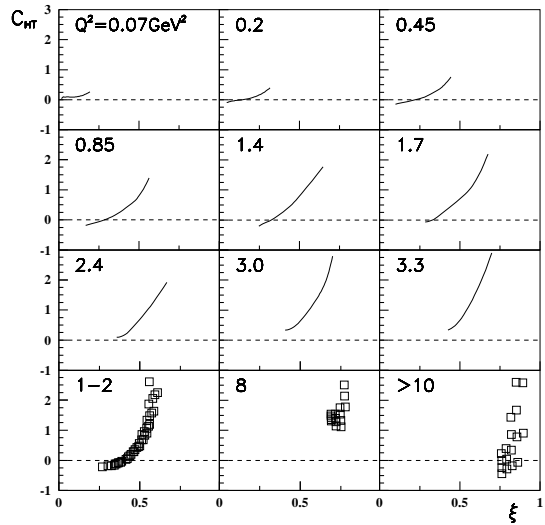}
\caption{
Higher twist coefficients $C_{HT}(\xi)$ (in $GeV^2$ units)
for inelastic lepton scattering on proton target.
}
\label{fig:3}
\end{center}
\end{figure}

The original quark-hadron duality [2] suggests
$$\overline{F}_2^p(x,Q^2)=F_2^{LT}(x,Q^2), \eqno(5)$$
i.e., the higher twist contributions tend to be largely canceled by the
Bloom-Gilman average [2]. Clearly, the duality in Eq. (5) is
obviously violated from the plots at $Q^2<1.7GeV^2$ in Fig. 1.
A direct measurement without pQCD analysis of parton distributions
can not separate the higher twist corrections even at high $Q^2$.
It is not a sound argument that an integral average of the measured structure function
at low $Q^2$ can be identified to be a pure leading twist structure
function without any higher twist contributions. Therefore, we give
an alternative explanation of the quark-hadron duality shown in Eq. (3).
There are two kinds of correlations among the partons: the
short distance correlation and the long distance correlation.
They are related to the higher twist effects in DIS process
and the resonance structure of nucleon, respectively.
The integral average of the resonance curves cancels the long distance
correlations and the resulting structure functions are related
to parton distributions with only the short distance
corrections - the higher twist corrections to DGLAP equations.

We calculated $F_2^{ZRS}$ at the $LL(Q^2)$ approximation, where the high
order corrections of $\alpha_s$ are neglected. Our predicted
distributions are comparable with the data at $Q^2>1 GeV^2$ [5],
it implies that the high order effects of $\alpha^n_s$ are absorbed
into a free parameter $R$ in the non-linear terms of the evolution equation.

The factorization in Eq. (1) is not proved at low $Q^2$ and small
$x$, where the correlations between initial and final states
is not negligible. However, we can use some phenomenological model to
mimic its contribution. We will discuss this elsewhere.

In summary, using the JLab data measured at low $Q^2$,
we separate the contributions of parton distributions
from higher twist corrections to lepton-proton DIS process
in the resonance region. We get the parton distributions at low $Q^2$
using the input valence quark distributions at $0.064GeV^2$ based on
the modified DGLAP equation. The concept of valence quarks and valence quark distributions
are indispensable even at $Q^2<1GeV^2$ in order to understand the nucleon
structure quantitatively. The quark-hadron duality
should be modified as Eq. (3). The higher twist corrections can not be neglected
to study the quark-hadron duality.


\end{document}